# Secrets of Successful Science Projects

Amy Courtney and Michael Courtney
Michael_Courtney@alum.mit.edu
Amy_Courtney@post.harvard.edu

**Abstract**
Over the past several years, the authors have served as teachers, qualified scientists, mentors, and/or parents on dozens of science projects. These projects ranged from elementary school projects that can be completed in a weekend to high school and college freshmen projects that take a semester or year to complete and yield published scholarly papers and/or compete at the highest national and international levels. This article describes what we have observed to be important to success.

**Introduction**
The need or desire for a student to complete a science project can be a source of both anxiety and opportunity. (Bochinski, 1996) A well-executed science project builds good skills in science and math, project planning, logical thinking, setting and keeping milestones, and can strengthen preparation for college and beyond. (Bochinski, 1996) A poorly-executed science project can be a source of stress and disappointment. (Grote, 1995) Over the past decade, we have mentored dozens of science projects completed by students ranging from middle school to freshmen in college. Dozens of projects have been published as scholarly papers, and we've never had a project fail to place in a regional science fair. Several students have won all-expense paid trips to national and international competitions, and over 70% of projects competing in state science fairs won 1$^{st}$ place in their category, with over 90% placing 1$^{st}$ or 2$^{nd}$ in category.

The early sections of this paper (up through the sections discussing the hypothesis) can be productively applied to all levels of projects with a wide range of efforts and goals from the weekend project trying only to earn a good grade to projects taking a full year and aspiring to win at the highest levels. The later sections of the paper are geared more toward projects hoping to perform well at the state level and beyond and yield results that are suitable for publication.

We observe that the best outcomes tend to occur when students are given the option to participate or refrain in different competitions. In other words, where there might often be value in requiring student research for evaluation in an educational context, a combination of the subjective nature of judging, student anxiety, and other factors makes it counterproductive to force competition in students who are reluctant. "Success" must be defined in terms of good science and research rather than in outcomes in subjective competitive events. (Czerniak, 1996; also see Yasar, 2006) Our view is that it is the research experience, rather than the competition, that produces positive outcomes for the student. (But see also Sahin, 2013 for a different view.)

**Identify the Goal and Desired Venue**
ISEF, i-SWEEEP, Broadcom Masters, Google Science Fair, Stockholm Junior Water Prize, Young Naturalist Awards, school science fairs, Junior Science and Humanities Symposium, and others all provide opportunities for student science projects to receive feedback and be judged in competitive environments. Numerous publication opportunities provide further opportunities for high quality projects to receive feedback from scientists and for good papers to take a permanent place in the scholarly literature. Most students begin with a school or regional science fair that is either ISEF affiliated or feeds into an ISEF-affiliated or Broadcom Masters-affiliated fair, but there are many other options.

Projects assigned by schools usually define the goals and venue, which also determines the rules for participation. However, projects pursued independently of school assignments should include due consideration of the goals, with an appropriate venue selected accordingly. All of the venues mentioned above have corresponding web sites, and many have national and regional or state web sites and points of contact to help get you started.

**Tight Correspondence Between Hypothesis and Method**
The most important feature of a project for success in most competitions, symposia, and publication venues is a tight correspondence between a scientific hypothesis and the experimental method chosen to test the hypothesis. No matter how interesting a hypothesis may be, if a method cannot be developed and executed by the student to adequately test the stated hypothesis, the project will not be very good, and other hypotheses should be considered.

This overriding feature is so important that students we mentor often take weeks or months brainstorming, considering, and rejecting a wide array of possible hypotheses for consideration. There is a pretty good correlation between project success, and the number of hypotheses that a student decided not to pursue. The cycle works something like this: 1) Student has an idea for hypothesis and writes it down or articulates it clearly in discussion with teachers, parents, and/or mentors. 2) Student does background research to consider what work has been done before and consider possible experimental methods. 3) Student outlines a possible method in writing and discusses with parents, teachers, and/or mentors for feasibility and correspondence with the hypothesis. 4) A few iterations of these steps may be needed before settling on a plan to move forward, but most hypotheses are rejected as a) already done b) not feasible with available resources or time c) not sufficiently interesting given the compromises that need to be made to test it.

Forming a hypothesis and determining a method to test it almost always requires a careful answer to two key questions: 1) What is the independent variable (often the hypothetical cause or a proxy for the hypothetical cause that can be observed and measured quantitatively)? 2) What is the dependent variable (often the hypothetical effect or a suitable proxy that can be measured and quantified)?

In our experience, a poorly considered hypothesis and test method is a main reason for poorly executed projects we see at science fairs or those we review for possible publication. When this occurs, it is not possible to salvage the scientific quality after the experiment is complete.

**Tight Correspondence Between Hypothesis and Results**
After an experiment is completed, there are often many possibilities for presenting the experimental data and results. Everyone likes some combination of tables, charts, and graphs. But students and parents need to be mindful that judges and peer reviewers are looking for simple and clear connections between the articulated hypothesis and how the results are presented.

Just like hypothesis selection and method development, deciding how to best present results is often an iterative process. Mentors and teachers should review the first attempts of the student and make suggestions for improvement. Would a bar graph be better than a table? Are error bars needed? Is the line graph the best presentation? Are all the axes and column headings clearly labeled with the quantity and units? Is the dependent variable on the vertical axis where judges are expecting it? Are the statistics done correctly with proper reporting of means, standard errors, P-values, and r-values? Is there enough data to support an interpretation or conclusion? Is the analysis done correctly?

Of all the competitive projects we've mentored, the worst a student ever finished was fourth place in a regional fair. The student had made significant mistakes in his analysis and refused to fix the mistakes when the mentor repeatedly pointed them out.

**Tight Correspondence Between Hypothesis and Discussion or Conclusion**
Projects with hypotheses that are unsupported, disproven, or only partially supported can be just as successful as projects with hypotheses that are strongly supported. But judges and peer reviewers are looking carefully and will quickly spot errors in discussions and conclusions that are not well founded in the data and analysis presented in the results section. Our experience is that young scientists need a lot of support from parents, teachers, and mentors in this area. Most likely, this is because students dearly want their hypothesis to be proven true and often make overly certain or overly optimistic word choices attempting to "sell" their ideas to their audience.

An essential trait acquired by a maturing scientist or student is the ability to communicate one's conclusions with the appropriate words and expressions of confidence.

**A Mentor**
It is widely accepted that students of science executing original research are best served by working under a knowledgeable mentor through graduate school and even in early post-doctoral work. Therefore, it seems reasonable that middle schoolers, high schoolers, and college undergraduates will also maximize their potential with guidance. The paperwork for ISEF affiliated fairs calls this individual the "qualified scientist" overseeing the project, and this designation may be the same or different from the individual listed as the "teacher" and also the "adult supervisor."

The central idea here is that "many advisers make victory sure." If neither the parent nor the teacher feels suitably qualified to serve as the mentor (qualified scientist), then outside expertise may need to be recruited. Mentors offer many advantages such as helping choose a hypothesis, developing a method, finding good references, ensuring the analysis is done correctly, and cautioning against incorrect or overly confident conclusions. Mentors are also likely to point out when projects have been done already, and are the best bet for good advice when assessing the likely cost and difficulty of a chosen topic. (See also DeClue et al., 2000)

A good mentor is essential for figuring out how to execute a given idea with the available resources or adapt a good idea to one that can be approached with the time and budget a student may have. With a bit of brainstorming with a mentor, many experiments can be realized with materials around the house, easily purchased at a neighborhood store, or easily borrowed from many high school science laboratories. Mentors will also be quick to realize how public databases may be used to acquire large volumes of quality data for testing interesting hypotheses without the challenges and (often) prohibitive expenses of performing original experiments. Some librarians are also beginning to specialize in helping track down data for projects. (Garritano and Carlson, 2009) Working with a mentor can also be useful to identify and make successful requests of non-public data that may be available from other researchers or government agencies for student analysis.

Mentors need not have a PhD or be overly familiar with a specific field. An undergraduate degree in science or engineering is sufficient if the mentor is willing to do some independent work familiarizing themselves with the areas of student interest to provide good support. Our degrees are in Physics and Engineering, but we have mentored successful projects in animal science, ecology, mechanical engineering, chemistry, mathematics, consumer product testing, blast injury, ballistics, and physics.

There are trade-offs between picking a local mentor who can meet and spend time in person with the student and a distant mentor who may be more qualified and more willing to spend time assisting the student, but whose contact may be limited to electronic communications. We've mentored a number of science projects from a distance, and this arrangement requires either strong local accountability or a very well motivated student who can make a timeline of milestones and deliverables and work hard to stick to it.  Skype, email, YouTube, and other social media provide excellent opportunities for high quality distance interactions if all the parties are willing to make good use of the available resources.

Distance mentoring requires either strong self-motivation to stick to a timeline, or a local authority (parent or teacher) holding the student accountable for timely completion of the many tasks required for successful project completion.  Another major reason that projects fail to meet student aspirations (in addition to inadequate hypothesis formulation or test design) is missed timeline milestones for defining and then executing the experimental portion.  Our experience is that if a good experiment is not completed several months before the final work product is due, there is rarely ample time to do a good job on analysis, presentation, and framing the results in the context of the literature and prior work.  The best projects have three to six months of additional effort after all the data is collected.

**Interesting Topic**
Our students often brainstorm and consider possible topics for 4-12 weeks.  Considering a topic means counting the cost and evaluating possible hypotheses and experimental methods.  Considering a topic means research and background and looking up 5-15 references to determine if similar experiments have been published before, how they turned out, and what the current project offers that might be new or different. (Also see LaBanca, 2008.)  For this reason, it is best for a student to begin this process the semester or summer before the project is actually begun - that is, before the timeline for accomplishment milestones even begins.

However, a great topic or catchy title is no substitute for tight correspondences between hypothesis, method, results, and discussion.  Tight correspondence is the core of the scientific method, and failure to execute the scientific method properly cannot be rescued with a clever topic or brilliant presentation.  Most evaluation rubrics focus on proper application of the scientific method.  (Note, the evaluation rubrics for specific science fair events are usually available online and should be kept in mind throughout the process.)

A topic needs to be interesting first of all to the student, because the level of interest will drive student motivation, enthusiasm, and effort.  Secondly, the topic needs to be interesting to the judges and reviewers who are evaluating the work in the chosen category.  Judges in the Physics, Engineering, Chemistry, and Math categories tend to find different things interesting, so consideration of the category or publication venue needs careful attention when gauging potential project interest.  Ecology, energy, and environmental topics are often trendy, so a novel idea executed well can draw great interest.  Topics of military relevance may draw much more interest in an area dominated by military influence but will be viewed as less relevant if the fair and its judges are associated with institutions that underappreciate the importance of the military.

If the location of the competition or publication venue is known, it is worthwhile perusing abstracts and titles of accepted papers and winning projects from previous years to determine what the judges of those venues tended to find interesting in the past in a given category.

**Novel Topic**
Relevance to both the general public and to other scientists will always depend on whether the work has been done before and whether the outcome is well known and established with a close connection to data.  Note, there is an important distinction between a result that is widely believed based on a theoretical argument and a result about which the experimental support is also well known.  Famous results can be repeated by the student, because repeatability is an essential element of the scientific method, but we would steer away from projects or experiments that have been done dozens of times. Success when repeating a well known result will often depend on approaching the experiment from a novel angle, or perhaps executing a clever, conceptually simple, inexpensive, or elegant experiment compared with the original experiments. Novelty is much more important at the senior level (high school) than the junior level (8$^{th}$ grade and under in the US.)

By novel, we mean that the same or similar hypothesis has never been tested before. Even if executed perfectly, topics that are regularly seen at science fairs usually have a top end of 3$^{rd}$ or 4$^{th}$ place at the regional level and have little hope of publication or winning at the state level. But take care to distinguish the specific hypothesis from a broad area. Over the years, there have been thousands of science projects on rocketry and ballistics, because students love shooting things and launching rockets. But to succeed in these areas, one really needs a novel hypothesis that the judges have never seen and even better if the experimental outcome cannot be determined in the first few minutes of a good internet search. Generally speaking, one should not expect a good outcome at the senior level if choosing a project topic from a list or book where the answer to the experimental question is already known. (Books like Bardhan-Quallen, 2007 are not recommended.)

**Volume of Work**
A project that can be completed in a weekend or two may well earn an A from a classroom teacher if it adheres to the tight correspondence between hypothesis, method, results, and conclusion discussed above, and it may (depending on the competition in a category) finish high enough in a regional fair to gain entry into a state level fair. But the volume of work possible in 10-20 hours of total effort is unlikely to be competitive at the state level, unlikely to yield a publishable work product (unless it is extremely clever) and unlikely to progress to national or international competitions.

Most projects that do well at the state level represent 100-200 hours of high quality student effort. This may seem like a lot until you compare it to the annual level of effort needed to succeed at the state level in most sporting competitions, music competitions, and other academic events. How many athletes spend hundreds of hours in practice each year and never even reach the quarter finals in their chosen field? This level of effort in a science project gives a much better chance at a high level of recognition and/or a published paper. This is best achieved with regular effort over time (5-10 hours per week depending on the phase of the project), which also results in a greater depth of understanding by the student.

Many private schools, charter schools, and magnet type schools provide structure for this level of effort over time by offering special science research courses, so student schedules permit a dedicated hour each day of the 180 day school year for the students' research projects. The most successful home school projects have provided this kind of scheduling and accountability as well.

Well-structured programs require timelines with documented (and graded) milestones such as the project topic, written hypothesis, background research, written method, acquisition of materials, execution of experiment, data collection, completion of fair paperwork, data analysis, full written paper, and review at each stage by an applicable combination of the teacher, parent (or adult supervisor), and mentor (or qualified scientist).

**Volume of Data – Accuracy of Data**
Just as a large bouquet of roses or a well written love song can make a lover weak in the knees, knowledgeable science fair judges can be impressed by a large volume of quality data that represents a valid test of an interesting hypothesis. A good fair presentation also takes due care to reveal the volume of work, but in a way that communicates sound science without screaming, "look how hard I worked."

Brainstorming ways to obtain over 1000 data points can really pay off, as can brainstorming ways to reduce the uncertainties to under 1%. Simple statistics suggest if 10 data points give an estimated error of 10%, you need 100 data points to reduce the error to 3% and 1000 data points to reduce the estimated error to 1%. The sources and magnitudes of the uncertainties, and the likely uncertainty of an experimental design are key aspects of designing a good experiment. This is an area where good input from a mentor is indispensable. There are trade-offs between improved experimental measurement technique for each data point, and reducing uncertainties in the mean with a large number of trials. Experience in skillful use and calibrating inexpensive equipment or choosing an independent variable or proxy that can be easily measured with much greater accuracy can be the difference between needing 10 data points and needing 1000 data points to reduce the error to 1% or so.

Most mentors will know that time and frequency are relatively easy to measure to 1% or better with inexpensive equipment available to most students. Measuring most other quantities this accurately requires greater experimental care and instrumentation calibration if it is possible at all. A mentor can also help design a simple pilot study and brief the student on standard analysis techniques to determine what sample sizes and tweaks to the experimental method may be needed before the full study. Publicly available data sets, as mentioned above, are another way to obtain adequate volumes of data for highly accurate analysis.

**Pilot Study**
A pilot study is a first attempt at executing the experimental design with a limited number of trials and/or a small sample size. The purpose is to identify any flaws in the design, estimate the uncertainties for a given sample size, and determine how much experimental effort will be needed for a larger number of trials. It provides key information to the scientist regarding whether she can decrease uncertainties simply by increasing the number of trials, or if she needs to significantly improve measurement accuracy of each trial. It also tells the scientist whether various aspects of the design need to be adjusted in order to run smoothly, control confounding factors, or legitimately test the desired hypothesis. Proper operation of any equipment should be confirmed at the same time.

We see a lot of projects at competitive fairs that could have been great had the original project been treated as a pilot study after a few trials. The experiment could have then improved after the flaws in the initial design became apparent by analyzing the data from the first few trials. The problem was that the students either failed to recognize that they were headed down a blind alley or were too lazy to improve their projects after experimentation had begin. Some projects we mentor have intentional pilot stages. Others simply realize after 5-10 trials that things are not working well enough to achieve the desired uncertainties and a valid test of the hypothesis, but with a few tweaks, one could have a much improved experimental design. It's OK to circle back and adjust the experimental design to account for things one learns in a first attempt and then begin the experiment anew with a blank slate. Further, judges will not mind and may be impressed to know that a pilot study was done.

**Well Prepared Display of Data and Results**
In addition to having a tight correlation between how the results are displayed and the hypothesis the experiment was designed to test, the display of data and results needs to be carefully considered and appealing. There should be uniformity in the style and color scheme. Fonts matter. Vertical and horizontal axis labels matter. Units matter. Line styles and thicknesses matter.

We typically have students prepare 5-10 times the graphs, tables, and charts for their research binders than actually end up on the display or in the paper. The data needs to be considered from many different perspectives before deciding which presentation most effectively communicates the results in a way that the audience can simply connect the results to the hypothesis as a bridge to the conclusions and discussion. Again for emphasis: presentation of results should serve as a well communicated bridge between the hypothesis and following discussion and conclusions.

Parents, adult supervisor, teacher, and mentor should all have valuable input here. Maybe the error bars need to have their line widths made more uniform, or the color scheme needs to be corrected, or the units got left off of one of the graphs, or in some results an abbreviation is used and in another the word is spelled out entirely, or the font size or style changes between graphs. Students often need to be reminded that older science fair judges lack the keen eyesight of teenagers, and the results ought not represent a seeing eye chart from where the viewers are likely to stand.

We are looking for the few graphs or tables or charts we like to call the "money shot" where the audience is likely to see for themselves whether or not the hypothesis is supported without needing a lengthy explanation to "connect the dots."

**Recognized Weaknesses and Limitations**
Enthusiasm is great, but even the best research projects have weaknesses and limitations. Odds are most judges and peer reviewers will spot some of them and are wondering if the student knows about them. A mentor and other professional input can be really helpful here.

Several questions should be considered: is the data really accurate enough to address the hypothesis? Could it be improved? Is there enough data? How broad are the implications? Would similar experiments give the same results, or is the answer likely to change with minor tweaks in the experimental design? What is the expected scope of this result? The student should practice simple, clear answers to these questions. The student should not feel bad when a judge asks questions like these. As long as the work was well conceived and executed, understanding the limitations of the project does not take away from the student's or the project's credibility.

It is important for the student to avoid grandeur and overly broad conclusions. The results might not be the same in other systems that are similar in some ways but different in others.

**Simple Enough to Do Experiment at Home**
Some may think the key to science fair success is an invitation to conduct research in a university laboratory with all the bells and whistles. Certainly a lot of science projects benefit from the equipment and opportunities afforded by well equipped laboratories. But our students have often found themselves in close competition with these projects, and we've come to think the judges and peer reviewers appreciate the initiative and creativity of doing high quality original research with the more limited equipment and resources available at home with a fraction of the budgets of well-funded university or corporate labs. Whether the student works in an established lab or independently, judges are interested in what the student has actually done, rather than in what others have done.

**Potential Future Work**
Successful science always considers what may be next given the implications of the current study. Perhaps the new method could be used more productively with a few tweaks. Perhaps a significant finding requires a larger sample size. Perhaps the scope and broadness of applicability of the result needs to be investigated to gain more confidence. Perhaps there is already a clear and compelling product or policy implication.

We see a lot of project presentations with an area on the board titled "Application to Society." This can be important for some projects, but trying to apply results from astrophysics to society can seem like a stretch. A better approach may be to understate applications of the current result, but use potential applications to motivate more careful future work. Conveying humility and the tentativeness of science is important, and it is hard not to seem like pride if a student suggests their project should compel major changes in industry or society or has really solved a very important problem.

**Write a Paper**
Students often think they have completely mastered all the nuances of thought, logic, and importance of their project. However, writing a complete scientific paper (Introduction, Method, Results, Conclusion) forces them to articulate their thoughts and logic in an intelligible manner. Often the first draft parrots a few key ideas repeatedly without a well developed logical flow.

Writing an introduction forces students to do their background research and distinguish relevant prior work from unrelated prior work. In considering prior work, students should think critically about the strengths that they might emulate and limitations they might improve upon in their own work. It provides an opportunity to articulate both the context and importance of the study. Both the thoughts that went into formulating their hypothesis and the hypothesis itself need to be clearly stated.

The logic and order of a method section are very important. All the reasons may not be included, but need to be considered and reviewed as the method section is written so that the student is prepared to answer questions during their presentation. The standard of detail desired in a method section is that another scientist in the field could repeat the experiment from the information provided and obtain the same results. Often figures or pictures showing the experimental apparatus or set-up are key in communicating the method.

The results section should present the data in an organized manner that allows the reader to connect the hypothesis from the introduction with the discussion or conclusions that follow. We encourage students to prepare the key charts, graphs, and tables first and think about what order to present them in the results section. Then a caption should be written describing each figure, table, or chart. The text of the results section should primarily serve to narrate through the figures, tables, and charts, with at least one paragraph describing the results presented in each.

There should usually be at least one paragraph in a discussion or conclusion section of a paper for each bullet point in the discussion or conclusion section of a presentation board. This helps the student provide a logical order to their discussion, expanding on the most important points first (the direct answer to the hypothesis) and then covering less important points (limitations, confounding factors, future work, applications, new hypotheses).

Some competitive events (SJWP, JSHS, etc.) require a written paper. Judges in presentation style events often like to see a written paper in the research binder, and there is no substitute for writing a paper to help students choose and practice and refine words and phrases to accurately describe their work to a scientific audience. A written paper can also be circulated among experts, teachers, and others to solicit feedback and questions in preparation for a competitive presentation.

**Worthy of Publication**
A written paper also gives students, mentors, teachers, and parents an opportunity to consider whether the work product is worthy of publication. It's also easy to get feedback from other scholars in the field whether and which venue might be appropriate for publication. Even if the paper is not publishable in the present form, scholars in the field can usually advise whether the underlying science project contains a publishable result if the paper is improved.

We have mentored dozens of science projects that have yielded published papers, and it is not uncommon that publication (rather than competition or presentation) is the desired goal of the project from the beginning. Publication of a good project is a more likely outcome than uncertain outcomes of a single high stakes presentation event with judges of unknown expertise and quality. We've seen the Physics category judged by dentists and high school students on occasion. We've never seen a paper we deemed worthy of publication fail to get published.

We always remind students in competitive events that judging science projects is an inherently subjective undertaking, and once their teacher, mentor, parents, and selves are satisfied, that they should not be discouraged if some judges think another student's project is better. Good science should be its own reward. Publication provides the opportunity to secure that reward regardless of what subjective judges do on a given day. Publication also puts the project results in the permanent literature for the benefit of scientists and students considering similar work in the future. If the peer-reviewers of one journal subjectively determine a paper is unworthy, it can be submitted to alternate scholarly venues. ISEF rules prohibit competing in more than one affiliated science fair.

**Practice Presenting**
There is no substitute for practice to help students weather the butterflies that appear in the stomach on science fair day. Students also benefit from feedback regarding how they really sound and how they think they sound. Practice presenting to parents, teachers, and mentors is a given. Presenting for other students can help get them out of their comfort zone and be forced to explain to an audience who may be less familiar with the project. I would also suggest that students participate in either the JSHS or make a video presentation (or both).

Our students have done very well in ISEF affiliated fairs, Broadcom Masters, and SJWP, but have never even made it past the first presentation stage of a state level JSHS (Junior Science and Humanities Symposium). Neither our colleagues nor students can figure out any reasons why, since on paper, the judging criteria are fairly similar. However, we think the JSHS is a great practice run for the ISEF-affiliated fairs for several reasons: 1) They require a written paper to be completed well in advance. 2) They require a 12 minute presentation in front of a room full of students, teachers, and judges. 3) The judges are almost certain to ask questions that the student has not already considered.

Making a 15 minute video of the student giving a power point type presentation is also great practice. Nothing reveals to a student their um, er, pregnant pauses, cliches, repeated phrases, and awkward facial expressions like video. Parents, mentors, and teachers need to take care not to be too critical of all these things. The student will likely see them for himself and work to improve. Usually a first video isn't good enough to send to subject matter experts for their feedback, but with a bit of practice, it is easy enough to produce a second video to distribute for better feedback from a tougher audience. Real scientists may not spend an hour or two reading a student paper and writing up feedback like a peer-reviewer, but they may watch a 10-15 minute video and fire off some quick thoughts and questions in a reply email.

**Mastery of Material and Background**
The higher a project goes in competition (school, regional, state, ISEF finals), the better the judges are, and the more likely they are to test students with hard questions about the material and background. Students who have really read all of their cited references multiple times and looked up uncertain vocabulary and concepts will be well-prepared; whereas, students who were faking it with ginned up confidence and fancy words will crash and burn. It's not too hard to buffalo judges at the school and regional level. State science fairs often have faculty experts in the various fields serving as judges, and the national and international events are guaranteed to be judged by experts with terminal degrees who are often chosen specifically to judge certain projects, because their expertise aligns closely with the project topic.

**Professional Appearance and Demeanor**
Some students do win wearing clothing that is hardly appropriate for school (much less scientific presentation). Since we acknowledge the inherently subjective nature of scientific competition, we should encourage students to give themselves the best opportunity by presenting a professional appearance and demeanor. Appropriate dress will also help the student feel confident in a stressful environment in which they may feel particularly self-conscious. We like to see young men in suits (or coat and tie) and young ladies in comparable business attire. If this is undesirable or unobtainable, a button up shirt with collar, dress slacks (at least not jeans), and matched shoes are far superior to jeans and a t-shirt (or worse). Both young men and young ladies should have the advice of an adult professional regarding appearance on fair day.

Certainly, every competitor should greet their judges with a firm handshake and look them in the eye while speaking, and especially when listening to and answering questions unless pointing something out on the board or referring to materials in the discussion process. Many students need some help and encouragement learning not to stare at the ground or off into space when speaking to adults. Students should treat judges like they want them to stay as long as possible and learn about their project, not like they are trying to shorten the conversation.

Don't let your students be the ones napping or who are glued to their cell phones, mobile devices, or video games while waiting for the next judge. Students should be reviewing their binder materials, reading an (even slightly) educational book, or productively engaged in something giving a scholarly and professional appearance when the judge rounds the corner and first sees them. Treat judges like they're the most important thing in the room as soon as they turn the corner, even if you think all the judges have been by for the day or they stop at another student project first.

**Rewards**
When middle school or high school athletes win first or second place in the state in a given sport, their accomplishments are likely to be memorialized for decades with a prominent banner in the school gym and a huge trophy in a glass case proudly displayed in the school foyer. Coaches are often given raises and contract extensions. The event is covered in local and state papers, and pages in the school yearbook are devoted to the accomplishment. Winning first or second place in a category at the state science fair is lucky to get a mention during the morning announcements. Is it any wonder we are producing generations who long for stardom in athletics or entertainment, but are struggling to produce enough STEM graduates to keep our workforce supplied and economy growing?

The goal of student science projects is not only to impart to students the ability to do good science, but a passion and love and confidence for doing it that will reap productive benefits for their whole lives, regardless of whether they become scientists or engineers themselves or contribute to the scientific education of following generations. This goal is best served with liberal and ample rewards and positive feedback for all levels of success.

We once mentored a sixth grader in a fisheries project who won the grand champion award at his regional science fair. Since no one else seemed like they would do it, one co-author (MC) lifted him to his shoulders and carried him around the room like he had just won the Super Bowl.

Different fairs have different kinds of awards: trophies, certificates, plaques, cash awards, computers, electronics, etc. We need to be sure we are making an effort at positive reinforcement so that students truly realize the nature of their scientific accomplishments. We recall in graduate school, our advisers always took all the students out for a special lunch or dinner to celebrate a paper being accepted for publication. Each student, school, mentor, and family will have their own ways of recognizing significant accomplishments, so our purpose here is more to inspire thought and suggest some possibilities.

If coverage in the school yearbook or newspaper is a possibility, this certainly should be encouraged. Even families of modest means should be able to manage a celebratory dinner or cake. A few dollars invested in a plaque or trophy or banner for a display case or school classroom would likely be money well spent encouraging students in future years that efforts and achievements will be remembered. We're sure with some thought and effort, teachers, mentors, parents, and school administrators can be sure to reward student accomplishments in a manner than encourages high levels of ongoing motivation and effort.